# Quantifying Oxidation Rates of Carbon Monoxide on a Pt/C Electrode

S. Balasubramanian, B. Lakshmanan,[1] C.E. Hetzke, V.A. Sethuraman,[2] J.W. Weidner[*]

Center for Electrochemical Engineering, Department of Chemical Engineering
University of South Carolina, Columbia, South Carolina 29208, USA

The electrochemical oxidation of carbon monoxide adsorbed ($CO_{ad}$) on platinum-on-carbon electrodes was studied *via* a methodology in which pre-adsorbed CO was partially oxidized by applying potentiostatic pulses for certain durations. The residual $CO_{ad}$ was analyzed using stripping voltammetry that involved the deconvolution of $CO_{ad}$ oxidation peaks of voltammograms to quantify the weakly and strongly bound species of $CO_{ad}$. The data obtained for various potentials and temperatures were fit to a model based on a nucleation and growth mechanism. The resulting fit produced potential- and temperature-dependent rate parameters that provided insight into the oxidation mechanism of the two $CO_{ad}$ species. Irrespective of the applied potential or temperature, the concentration of weakly bound $CO_{ad}$ species decreased exponentially with time. In contrast, the strongly bound $CO_{ad}$ species showed a gradual transition of mechanisms, from progressive nucleation at relatively low potentials to exponential decay at high potentials.

[1] – Present address: General Motors Fuel Cell Activities, 10 Carriage Street, Honeoye Falls, New York 14472, USA.
[2] – Present address: School of Engineering, Brown University, Providence, Rhode Island 02912, USA.
[*] – Corresponding author: Tel: +1 (803) 777-3207; Fax: +1 (803) 777-8265; E-mail: weidner@cec.sc.edu

## 1. INTRODUCTION

The need for carbon monoxide (CO) tolerant catalysts for proton-exchange-membrane fuel cells (PEMFC) has brought considerable attention to understand the kinetics of CO adsorption and oxidation on single and polycrystalline platinum (Pt) electrodes [1]. Various electrochemical and spectral techniques have been employed to: (i) delineate the mechanism of CO poisoning in PEMFC; (ii) estimate the CO induced performance loss in PEMFC; and (iii) develop techniques to mitigate the performance loss [2-7]. Electrochemical filtering is one such technique, in which the concentration of CO in the fuel stream is decreased by electrochemically oxidizing it to $CO_2$ [8-10]. Developing CO tolerant catalysts and designing electrochemical filters require a quantitative understanding of CO adsorption and electro-oxidation kinetics. In the literature, the kinetic parameters for the electro-oxidation of CO adsorbed on Pt were either measured from electrode-in-solution experiments [11] or deduced from fitting a model to the CO poisoning induced fuel cell performance-loss data [12-13]. However, *in situ* measurements of electro-oxidation kinetic parameters from electrodes reflecting PEMFC are rare due to the complex overlapping of different oxidation mechanisms.

The generally accepted mechanism for CO electro-oxidation is the Langmuir-Hinshelwood reaction between adsorbed CO ($CO_{ad}$) and a neighboring oxygen-containing species, which we refer as $OH^*_{ad}$, generated from hydrolysis (Eqn. 1) [14].

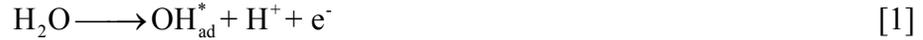
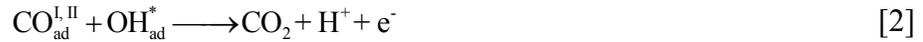

$$H_2O \longrightarrow OH^*_{ad} + H^+ + e^- \qquad [1]$$
$$CO^{I,II}_{ad} + OH^*_{ad} \longrightarrow CO_2 + H^+ + e^- \qquad [2]$$

The superscripts I and II indicate the two CO ad-species. It has been hypothesized that this reaction is limited by the availability of the $OH^*_{ad}$ adjoining a $CO_{ad}$ molecule [15]. Different models proposed to explain the accessibility of the adjacent $OH^*_{ad}$ include nucleation and growth (NG) of $OH^*_{ad}$ islands that grow to consume the $CO_{ad}$, surface diffusion of CO ad-species to active sites on which the CO oxidation reaction occurs [16-21]. NG mechanism treats the $CO_{ad}$ as stationary species on Pt surface; this approach explains the skewed potentiostatic current response of CO covered electrode [22-23]. The surface diffusion mechanism treats $CO_{ad}$ as mobile species diffusing towards active sites for reaction; this explains the symmetric CO oxidation peak during chronoamperometry and surface defects effects [17, 21, 24-28]. However, it was difficult to arrive at a physically justifiable yet experimentally quantifiable general solution from the models proposed for the electro-oxidation of $CO_{ad}$ on a Pt/C electrode. For example, Friedrich *et al.* [29] and Koper *et al.* [16] established the importance of the surface mobility of the species, and their single-crystal electrode studies showed higher activity in kinks rather than terraces supporting the active sites [30]. While Chang *et al.*'s observation of CO islands supports the NG mechanism [31], NG mechanism could not explain the role of crystalline defects in CO electro-oxidation [28]. As the experimental evidence agrees with each of the models under a limited set of conditions, a single physical model has not been shown to explain the CO electro-oxidation rates under most of the possible conditions.

In addition, $CO_{ad}$ stripping cyclic voltammetry (CO-SCV) on Pt electrodes shows a dual peak response during the electro-oxidation of $CO_{ad}$ [14, 32]. Gilman attributed this to the two prominent CO adsorption forms (ad-species) - atop and bridged [14], which were first observed by Eischens and Pliskin through infrared techniques [33]. Various explanations presented in the literature for the two CO ad-species include: particle size variation [27, 34-35], terrace *vs.* edge sites distribution [36], crystallographic orientation [24, 28, 37], difference in nucleating sites of oxygen-containing species [17] and difference in the mobility of surface species on different crystal facets [38]. The widely accepted view is that the differences in the pattern of adsorption on Pt exhibit the two oxidation peaks in a CO-SCV [5, 25, 39-50]. By deconvoluting the dual peak into individual peaks, representing the two ad-species, our group has quantified desorption and rearrangement kinetics of the two CO ad-species over polycrystalline platinum supported on carbon (Pt/C) electrodes [51].

The objective of this work is to develop and test a semi-empirical methodology to quantify the potentiostatic oxidation rates of the two distinct CO ad-species on a polycrystalline Pt/C electrode. This was done by quantifying the total unreacted $CO_{ad}$ after applying a potentiostatic pulse for certain duration on a Pt/C electrode with pre-adsorbed CO. The unreacted $CO_{ad}$ is quantified by CO-SCV. The dual peaks observed in the CO-SCV were deconvoluted into two Gaussian peaks representing the two CO ad-species. This procedure was repeated for different pulse durations under a constant potential to obtain the change in the overall CO coverage with pulse time. The electro-oxidation kinetic parameters of the two ad-species were deduced by fitting an equation derived from NG mechanism with the fractional unreacted $CO_{ad}$ for different pulse durations. The model predictions of the CO ad-species coverage were compared and validated with the individual ad-species coverage estimated from the deconvoluted CO-SCV. This procedure was repeated for different cell potentials and temperatures.

## 2. EXPERIMENTAL

*2.1. Experimental Setup*

The membrane electrode assembly (MEA) used has a Nafion 115 membrane sandwiched between two polycrystalline Pt/C (XC-72R, E-Tek) electrodes with a Pt loading of 0.5 mg/cm$^2$ coated over a carbon cloth gas with an area of 10cm$^2$. The MEA was assembled into a cell with single-channeled serpentine graphite flow fields. The electrodes were conditioned as reported in our previous work [51]. A gas concentration of 500 ppm of CO in nitrogen, flowing at the rate of 100 cm$^3$/min for 5 minutes was used as the CO source to saturate the working electrode. A gas mixture of 4% hydrogen in nitrogen was passed (100 cm$^3$/min) through the other electrode, which acted as both counter and reference electrode. All the gases used were procured from Air Products, Inc. and were certified for purity. The electrochemical experiments were conducted using a M263A potentiostat/galvanostat from Princeton Applied Research, Inc. and ECHEM software from EG&G.

*2.2. CO electro-oxidation and SCV*

To estimate the oxidation rate of $CO_{ad}$, a potentiostatic pulse was applied on a CO-saturated electrode for certain duration, $t_d$, at a constant temperature, followed by a CO-SCV. The SCV is carried out by scanning the electrode from 50 to 1,100 mV, with respect to reference electrode, back and forth for 3 cycles at a rate of 50 mV/s. The CO-SCV gives the quantity of unreacted CO after the pulse. The change in the quantity of the unreacted $CO_{ad}$ after different pulse durations indicates the oxidation rate of $CO_{ad}$ for that particular applied potential. This procedure of CO saturation and potentiostatic oxidation followed by stripping voltammetry is repeated for pulses of different durations, at different applied potentials in the range of 450~700 mV and for temperatures of 25, 45, 60 and 70°C.

*2.3. Analysis of a CO-SCV*

Figure 1 compares the anodic scans of CO-SCV (line) and a blank, *i.e.*, no CO adsorbed (dashed line). Compared to the blank one, the hydrogen oxidation region (< 400 mV) is depressed in the CO-SCV. This is due to the occupation of active sites by $CO_{ad}$. Whereas in the CO oxidation region (400 to 900 mV), the dual peaks in CO-SCV represent the electro-oxidation of the two CO ad-species. As the blank did not have $CO_{ad}$, there was no peak in that potential range. The net charge, corresponding to the area under the dual peak, includes the charge for CO electro-oxidation as in equation 1 and 2. To correct for the background current including double layer capacitance and Pt oxidation, the anodic current response of the blank was subtracted from CO-SCV. The area under the background corrected CO oxidation current is shown as the shaded region in the potential range from 400 to 900 mV. All of the CO oxidation peaks presented henceforth are background corrected CO oxidation peaks obtained from respective CO-SCV.

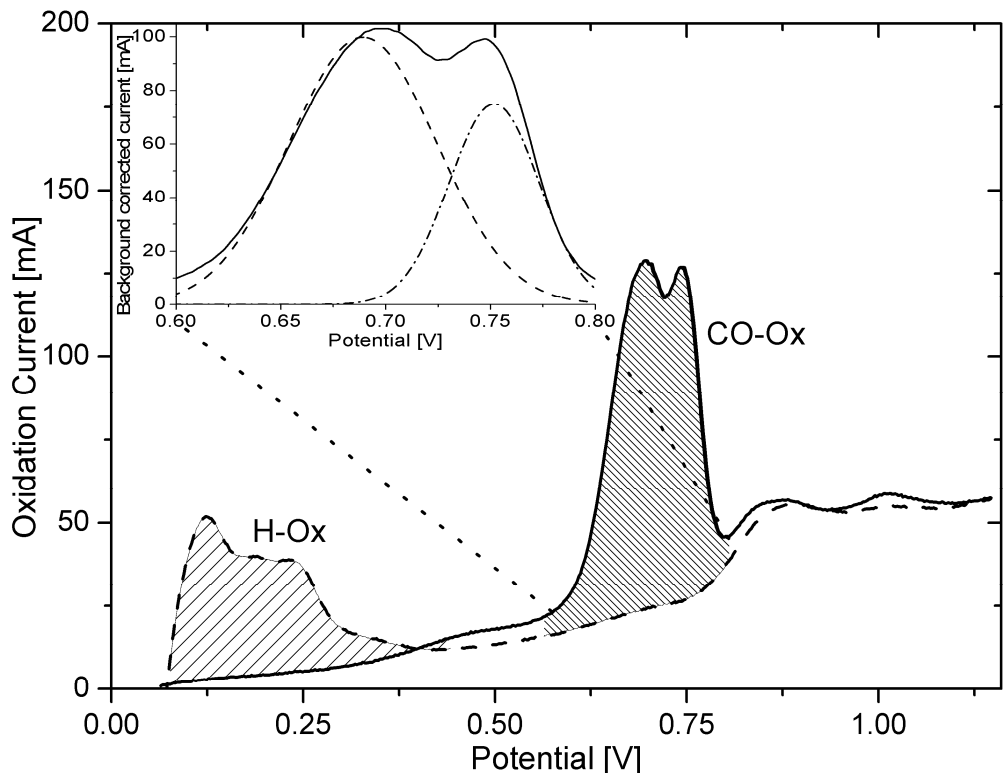

*Figure 1: Anodic scans of cyclic voltammograms obtained on a 40% Pt/C electrode at 45°C and atmospheric pressure. The dotted line represents Pt and the solid line represents Pt pre-adsorbed with CO. Inset shows the deconvolution of the CO oxidation peaks as two Gaussian peaks.*

*2.4. Deconvolution of the CO oxidation peaks*

The magnified portion of the background corrected CO oxidation peaks is shown in the inset of Fig. 1. The dual peak was deconvoluted as two individual Gaussian peaks representing the two ad-species, $CO_{ad}^{I}$ and $CO_{ad}^{II}$. For more information on the deconvolution of peaks, please refer to our previous work [51]. The peak at lower potential was assigned to $CO_{ad}^{I}$ and the peak at higher potential was assigned to $CO_{ad}^{II}$. The oxidation charge for each of the ad-species was estimated by integrating the background corrected and deconvoluted oxidation current peaks over time and was referred as $Q^I$ and $Q^{II}$, respectively. This charge is proportional to the quantity of the unreacted CO ad-species the Pt/C electrode (Refer eqn. 1 and 2). The fractional coverage of unreacted $CO_{ad}$ was estimated by taking the ratio of $CO_{ad}$ coverage after the application of a potential pulse with respect to the saturation $CO_{ad}$ coverage with no pulse applied. The fractional surface coverage of each of the CO ad-species ($\theta^I$ and $\theta^{II}$) is defined in the equations 3 and 4.

$$\theta^I = \frac{Q^I}{\left[Q^I + Q^{II}\right]_{t=0}} \quad [3]$$

$$\theta^{II} = \frac{Q^{II}}{\left[Q^I + Q^{II}\right]_{t=0}} \quad [4]$$

The total initial fractional coverage of CO saturated electrode, when no pulse applied, is taken as unity (i.e. $\theta^T_{t=0} = \theta^I_{t=0} + \theta^{II}_{t=0} = 1$). The change in the total quantity of $CO_{ad}$ with pulse time is taken as the fraction change in $CO_{ad}$ from the initial saturation $CO_{ad}$ coverage.

*2.5. Model Equation*

The electro-oxidation of CO is a surface reaction between CO ad-species and neighboring $OH_{ad}^*$ formed from the hydrolysis of water. To estimate the oxidation rates of the CO ad-species, the bimodal $CO_{ad}$ oxidation current peaks are treated as current response of two independent Gaussian populations that oxidize independently and simultaneously, albeit with differing rates. The desorption and re-arrangement rates, estimated in our previous work [51], are an order of magnitude lesser than the oxidation rates and are assumed to be negligible in relation to the oxidation rates.

According to NG mechanism, CO electro-oxidation rate increases with the rate of nucleation and growth of $OH_{ad}^*$ islands and then decreases as these growing $OH_{ad}^*$ islands overlap resulting in the shrinkage of $CO_{ad}$ islands [22]. Vollhardt and Retter lumped the nucleation, growth and collapse of islands to a simple expression of fractional coverage as a function of time as [52],

$$\theta_t = \theta_{t=0} \exp(-Kt^X) \quad [5]$$

Where K is the lumped rate constant and X is the order of time. X indicates the type of NG mechanism and takes the value 1, 2 or 3 for the limiting cases of exponential decay, instantaneous nucleation and progressive nucleation [51], respectively. Extending this relation for two independent populations of CO ad-species, the fractional coverage of CO can be written as,

$$\theta_t^T = \theta_{t=0}^I \exp(-K^I t^{X^I}) + \theta_{t=0}^{II} \exp(-K^{II} t^{X^{II}}) \qquad [6]$$

Love and Kapowski observed that part of the Pt crystal facets exhibit progressive nucleation and other part exhibit instantaneous nucleation, for the same applied potential [23]. Therefore, we consider $X^{I,II}$ as a parameter that encompasses the changing nature of oxidation mechanism from progressive nucleation to exponential decay mechanism for different crystal sites.

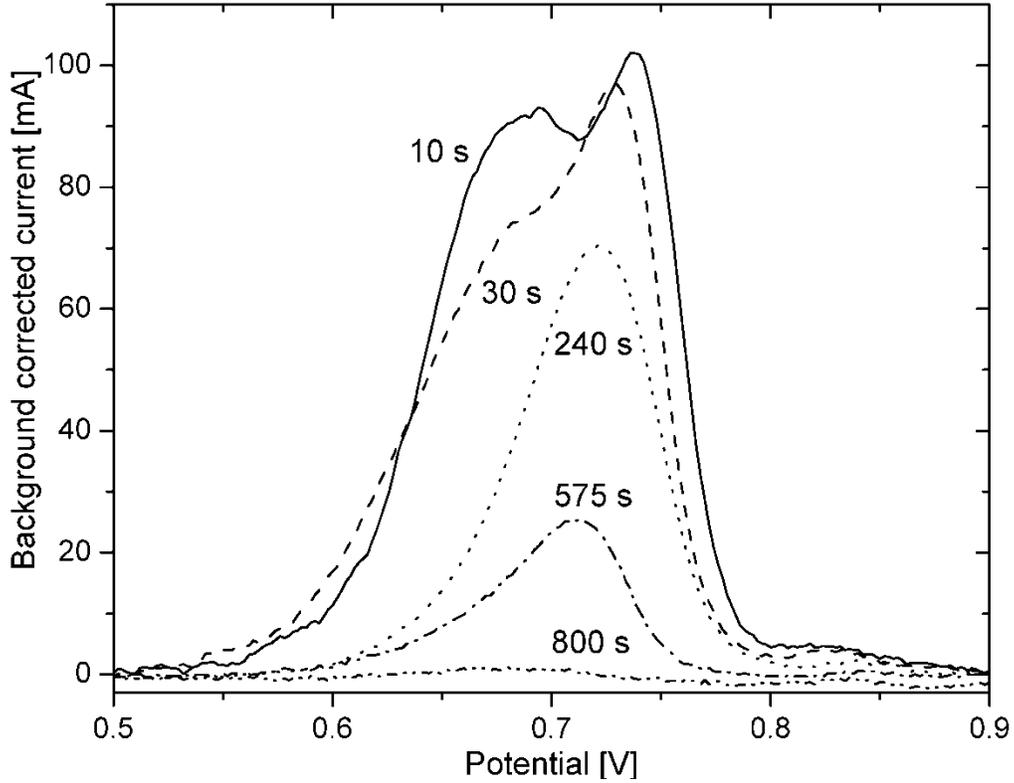

*Figure 2: CO oxidation peaks obtained from baseline corrected CO-SCVs on CO saturated Pt/C electrode after applying pulse potential 550 mV for different durations at 45°C.*

## 3. RESULTS AND DISCUSSION

*3.1. Effect of oxidation time*

Figure 2 shows the effect of pulse duration on the CO oxidation peaks obtained at 45°C for the potential 550 mV. With the increase in the duration of potential pulse, the height of the CO oxidation peaks decreased. The height of the peak 1 decreased rapidly than that of the peak 2. For the pulse duration of 240s, the peak 1 is absent indicating a complete oxidation of the $CO_{ad}^{I}$. This shows that $CO_{ad}^{II}$ oxidizes at a lower rate, which implies the significant difference in the oxidation rates of the two ad-species.

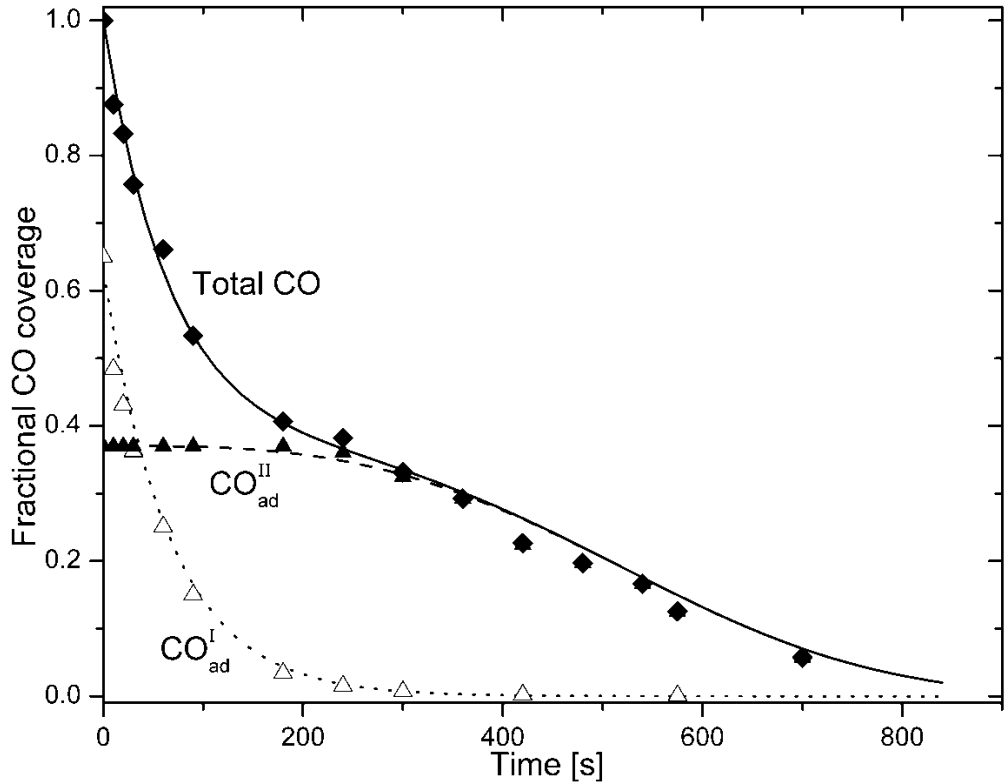

*Figure 3: Comparison of the predicted and experimental fractional CO coverage at 45°C and for pulse potential of 550 mV. The lines represent predictions and symbols show experimental results ( ◆-total CO, △- $CO_{ad}^{I}$, ▲ $CO_{ad}^{II}$ ).*

The integrated total charge under the $CO_{ad}$ oxidation peaks, obtained at 45°C for different durations of 550mV potential pulse, were fitted against the equation 6 and parameters $K^{I,II}$ and $X^{I,II}$ were estimated for the two ad-species. The fractional coverage of each of the CO ad-species was estimated by using the estimated parameters in the equation 5. Fig. 3 compares the experimentally measured and deconvoluted $CO_{ad}$ coverage (symbols) with the fitted and predicted $CO_{ad}$ coverage (lines) for different duration of potential pulses. The figure shows fractional coverage of CO ad-species predicted using the oxidation rate parameters, which were estimated from fitting the kinetic equation 6 with the overall coverage, agrees with the coverage calculated from the deconvolution of CO-SCVs. The total coverage declined rapidly during initial period followed by a gradual decline indicates two different oxidation rates. The sharp initial drop in the overall coverage agrees with the higher oxidation rate of the $CO_{ad}^{I}$ ad-species. This is consistent with the sharp decrease in the height of peak 1 with time, as observed in Fig. 2. The delay in oxidation of the $CO_{ad}^{II}$ ad-species was attributed to slow rate of initiation of the nucleation of the $OH_{ad}^{*}$ islands. This is evident from the fitted values of X for $CO_{ad}^{I}$ and $CO_{ad}^{II}$ ad-species, which are 1 and 2.3 corresponding to exponential decay and nucleation and growth mechanism.

*3.2. Effect of applied potential*

Figure 4 shows the base-line corrected CO oxidation peaks obtained after applying pulse potentials (550, 600, 650, 700 mV) for 10 seconds at 45°C of the CO saturated Pt electrode. This figure shows the effect of applied potential on the oxidation of $CO_{ad}$ from the electrode surface. With the increase in applied potential, the overall area under the peaks decreased at a faster rate. The first peak decreased rapidly and its contribution was not distinguishable for the CO peaks obtained for the pulse potentials 650 and 700 mV. In addition, as the applied potential was increased, the CO oxidation time decreased. This shows that an increase in the pulse potential promotes the CO oxidation reaction.

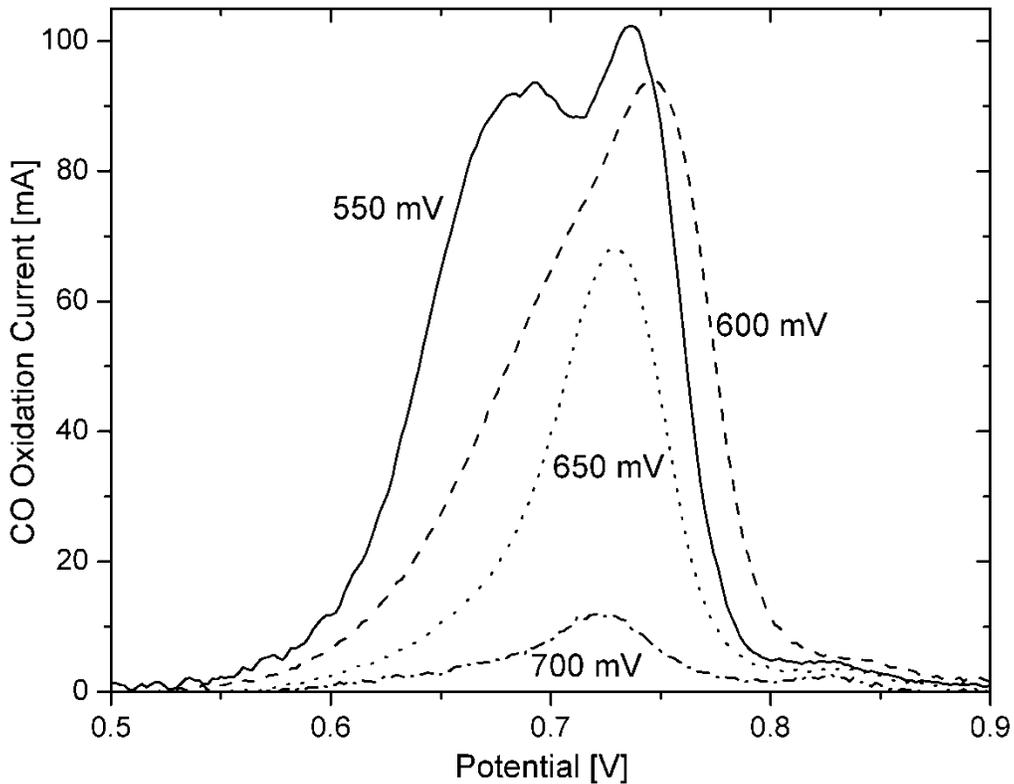

*Figure 4: CO oxidation peaks obtained from baseline corrected CO-SCVs on CO saturated Pt/C electrode after applying pulse potential (550, 600, 650, 700) mV for different durations at 45°C.*

Figure 5 shows the fitted fractional coverage change with time for various pulse-potential amplitudes (*i.e.*, 500 - 700 mV *vs.* DHE) at 45°C. The line represents the fitted results and the symbols represent the fractional CO coverage measured after applying the pulse. When 500 mV was applied, the CO coverage gradually dropped from around 1.0 to 0.25 after 800 seconds. However, when 700 mV was applied within 20 seconds the coverage dropped to zero. For intermediate cases of 550, 600 and 650 mV pulse, a sharp initial drop followed by a relatively stagnant period and a quick drop in the CO coverage is observed. At lower potentials, both of the CO ad-species oxidize at a very low rate. However as the potential was increased, nucleation of the $OH_{ad}^*$ was facilitated near $CO_{ad}^{I}$ ad-species contributing to the increased oxidation of $CO_{ad}^{I}$, while $CO_{ad}^{II}$ being relatively dormant until the nucleation of $OH_{ad}^*$ starts near the $CO_{ad}^{II}$ islands. The procedure of parameter estimation and deconvolution was repeated for the CO oxidation data obtained for temperatures 25, 60 and 70°C and the parameters of K and X were estimated for both of the CO ad-species.

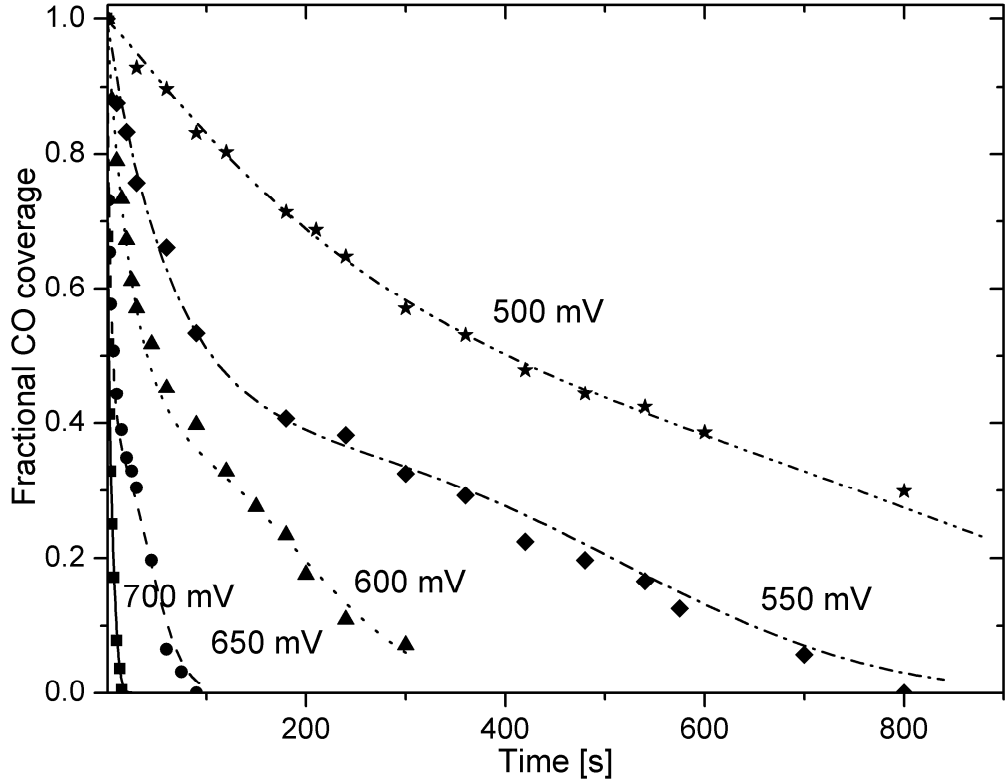

*Figure 5: Comparison of experimental fractional CO coverage (symbols) and fit from equation 8 (lines) at 45°C for pulse potentials of varying amplitudes between 500 and 700 mV vs. DHE.*

*3.3. Effect of temperature*

The effect of temperature on CO oxidation is shown in Fig. 6. The $CO_{ad}$ oxidation peaks shifted to lower potential with the increase in temperature from 25°C to 70°C. Also, the total peak area decreased with increase in the temperature, for e.g. when the temperature was increased from 25°C to 60°C, CO peak area dropped from 688 to 551 mC. The distance between the peak centers decreased with an increase in the temperature, i.e. from two distinguishable peaks at 25°C to the seemingly indistinguishable peak at 60°C. This shows that oxidation rates of the two CO ad-species are equivalent, at higher temperatures.

In summary, an increase in temperature, potential or pulse duration promotes the CO oxidation, which is observed from the decrease in the CO oxidation peak area after the pulse. It is also observed that the two peaks respond differently for a given applied potential. The distinct oxidation rates shows that the two peaks represent two independent populations of CO ad-species adsorbed on the Pt.

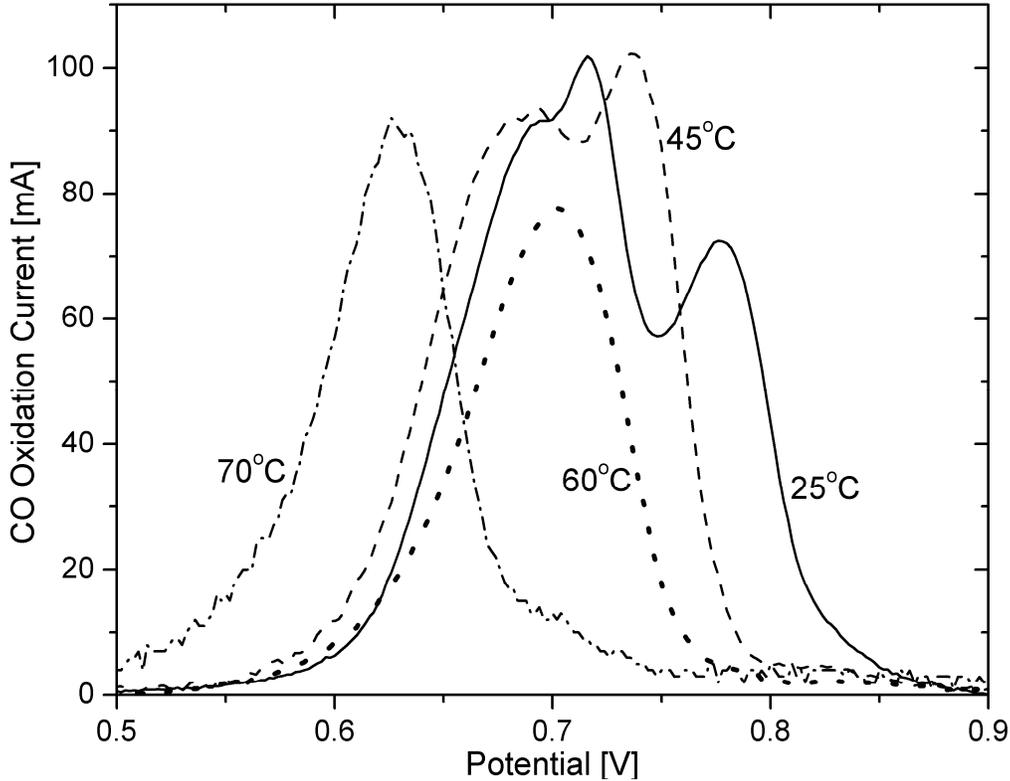

*Figure 6: CO oxidation peaks obtained from baseline corrected CO-SCVs on CO saturated Pt/C electrode after applying pulse potential of 550 mV for 10 seconds, for temperatures 25, 45, 60 and 70°C.*

Figures 7 and 8 show the potential dependence of the fitted parameters $K^I$ and $K^{II}$ for different temperatures (25, 45, 60 and 70°C), respectively. The figure shows that the rate constant increases with increase in temperature and oxidation potential, which is consistent with previous observations. It should also be noted that for the potential of 700 mV cases, the values $K^I$ and $K^{II}$ were closer indicating a comparable oxidation rate, which indicates a possible convergence of oxidation rates at high potentials.

Figure 9 shows the temperature and potential dependency of parameters $X^I$ (open) and $X^{II}$ (closed) corresponding to the two CO ad-species. The value of $X^I$ being closer to 1 suggests that the oxidation mechanism for $CO_{ad}^I$ population is exponential decay. It means for the potential window studied, the sites neighboring $CO_{ad}$ are available for $OH_{ad}$ and hence the oxidation rate is limited by the availability of $CO_{ad}$. Whereas, the value of $X^{II}$ decreased from values greater than 3 to 1 indicate a shift in oxidation mechanism from progressive nucleation to exponential decay with increase in the applied potential.

The shift in mechanism, with increase in temperature and potential, is attributed to the increase in availability of sites for $OH_{ad}$ nuclei formation, which may finally reach a maximum point at which all of the sites are available resulting in exponential decay mechanism. At this point, the rate of oxidation of strongly bound $CO_{ad}^{II}$, is expected to be limited by the reaction between $CO_{ad}^{II}$ and $OH_{ad}$, which explains the convergence of rates of the different ad-species at higher temperatures and potentials.

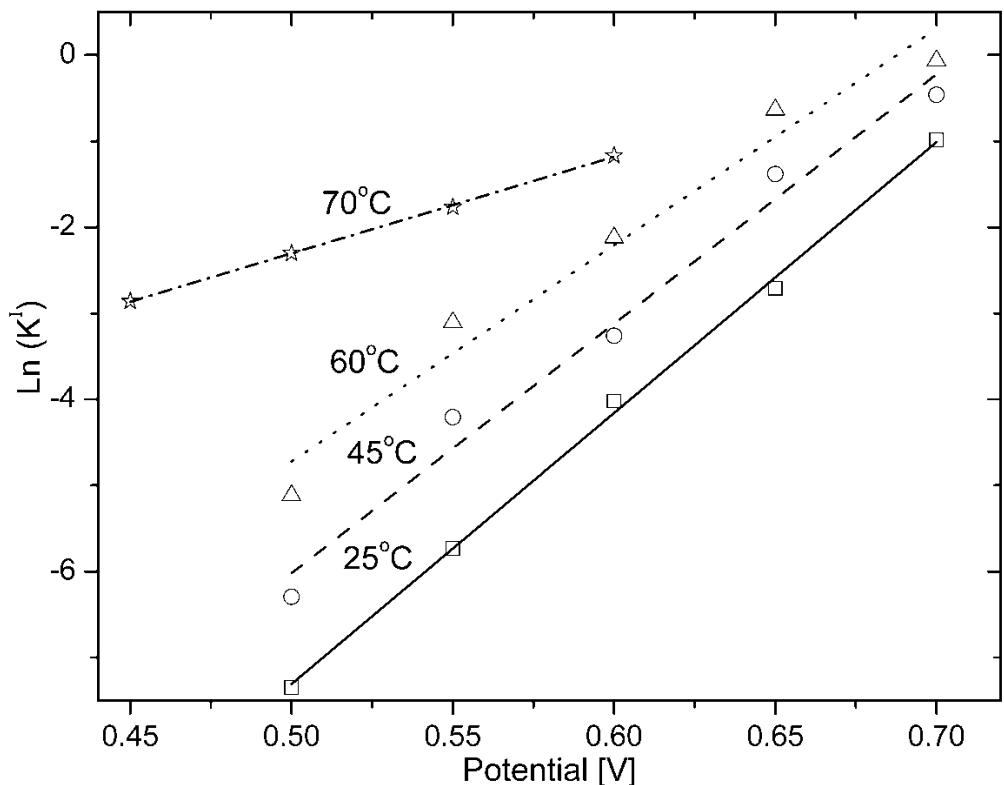

*Figure 7: Dependence of oxidation reaction rate constant ($K^I$) of $CO_{ad}^{I}$ ad-species on the potential for different temperatures (□-25, ○-45, △-60, ☆-70°C).*

## 4. CONCLUSIONS

A methodology was developed and tested to quantify the electro-oxidation rates of the two CO ad-species on polycrystalline Pt/C porous electrode. CO adsorbed was oxidized by systematic application of pulse potential for certain duration followed by cyclic voltammetry. The resulting change in the $CO_{ad}$ coverage was used to estimate the oxidation rate parameters by fitting against an equation based on nucleation and growth model. It was observed that each of the CO ad-species exhibits different oxidation mechanism. We attribute the difference to the formation and accessibility of $OH_{ad}$ to the different ad-species. Also, the oxidation mechanism of strongly bound ad-species changed from progressive nucleation to exponential decay. This explains the limitation of conventional approach to fit a single mechanism. The shift in mechanism is attributed to the increasing number of $OH_{ad}$ nuclei with increase in potential. The prediction of the proportion of the two ad-species obtained from estimated rate constants were found to match the deconvoluted voltammograms obtained after potentiostatic oxidation of CO. Using the technique described in this work, accounting for different CO ad-species and shifting mechanism, CO electro-oxidation behavior of a porous electrode could be measured *in situ*.

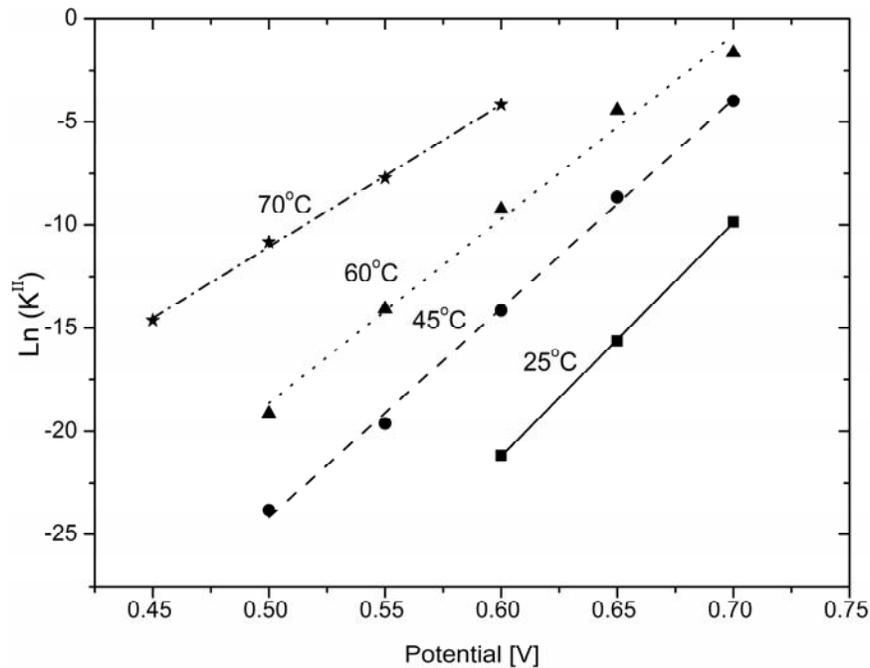

*Figure 8: Dependence of oxidation reaction rate constant ($K^{II}$) of $CO_{ad}^{II}$ ad-species on the potential for different temperatures ( ■-25, ●-45, ▲-60, ★-70°C).*

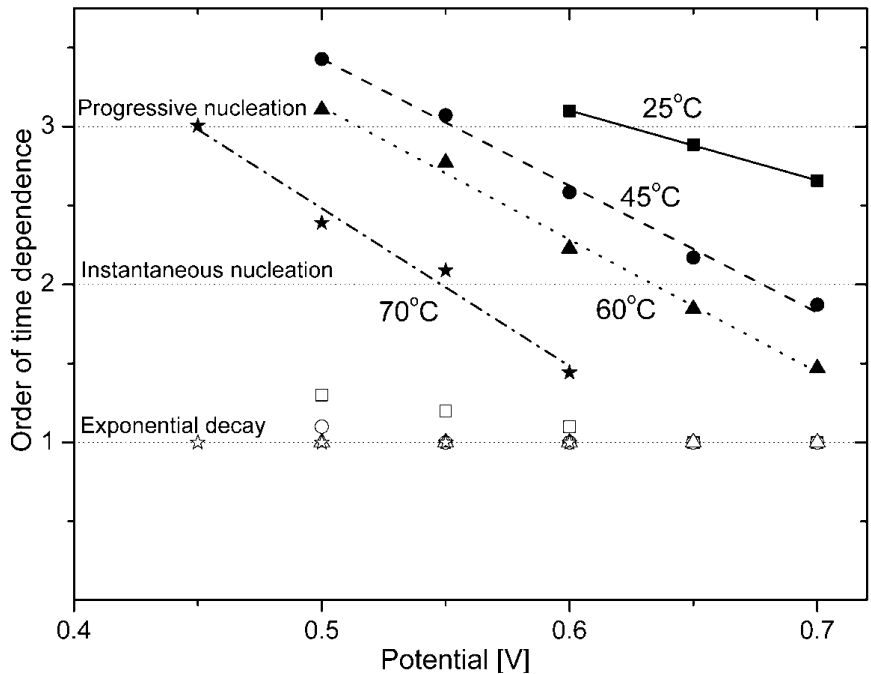

*Figure 9: Order-of-time dependence of $X^I$ and $X^{II}$ for $CO_{ad}^I$ (open symbols) and $CO_{ad}^{II}$ (closed symbols) ad-species, respectively, on the applied potential for different temperatures ( ■ - 25, ● - 45, ▲ - 60, ★ - 70°C).*

## 5. ACKNOWLEDGEMENTS

The authors gratefully acknowledge the support from the National Reconnaissance Office for Hybrid Advanced Power Sources under grant # NRO-00-C-1034 and the National Science Foundation Industry/University Co-operative Research Center for Fuel Cells under award # NSF-03-24260.

## 6. LIST OF SYMBOLS

Q - Coulombic charge
K - Lumped rate constant
X – order of nucleation

*Greek symbols*
θ - fractional coverage

*Super scripts*
I – pertaining to first peak
II – pertaining to second peak
T – total CO

*Sub scripts*
ad - adsorbed
CO – pertaining to CO
t - corresponding to time 't'